\documentclass{article}

\usepackage[utf8]{inputenc}		
\usepackage[english]{babel}		
\usepackage[T1]{fontenc}		

\usepackage[fleqn]{empheq}	       
\usepackage{mathptmx}                
\numberwithin{equation}{section}
\usepackage{amssymb}                 
\usepackage[fleqn]{cases}             
\usepackage{mathrsfs}                 
\usepackage{bm}                          
\allowdisplaybreaks

\DeclareSymbolFont{calletters}{OMS}{cmsy}{m}{n}
\DeclareSymbolFontAlphabet{\mathcal}{calletters}

\usepackage[nointegrals]{wasysym}
\usepackage{stmaryrd}
\usepackage{esint}			
\usepackage[misc]{ifsym}
\usepackage{extarrows}

\usepackage[margin=1cm, font={small,it}]{caption}
\usepackage[]{graphicx}		
\usepackage{grffile}
\usepackage{float}
\usepackage[section]{placeins}
\usepackage{subfig}
\usepackage{wrapfig}		
\usepackage{epstopdf}		
\graphicspath{ {./Figures/} }
\usepackage{xcolor}

\usepackage{enumitem}

\usepackage{array}			
\usepackage{dcolumn}    
\usepackage{booktabs}
\usepackage{tabu}
\usepackage{tabularx}
\usepackage{multirow}
\usepackage{diagbox}

\usepackage{tikz-cd}
\usetikzlibrary{shapes,arrows}
\usepackage{extarrows}

\usepackage{siunitx}

\usepackage{appendix}	

\usepackage[draft]{pdfcomment}

\usepackage{verbatimbox}
\usepackage{xspace}

\usepackage{authblk}

\newcolumntype{.}{D{.}{.}{-1}}
\newcolumntype{R}[1]{>{\raggedright\arraybackslash$\displaystyle}p{#1}<{$}}
\newcolumntype{L}[1]{>{\raggedleft\arraybackslash$\displaystyle}p{#1}<{$}}
\newcolumntype{K}[1]{>{\centering\arraybackslash$\displaystyle}p{#1}<{$}} 
\newcolumntype{C}{>{\centering\arraybackslash$\displaystyle}c<{$}}

\newcolumntype{G}[1]{>{\raggedright\arraybackslash }b{#1}}

\providecommand{\operatorname}[1]{\mathop{\mathrm{#1}}\nolimits}

\DeclareMathOperator{\ci}{\imath}
\renewcommand{\Re}{\operatorname{Re}}
\renewcommand{\Im}{\operatorname{Im}}

\newcommand{\diff}{\mathop{}\mathopen{}\mathrm{d}}
\newcommand{\derive}[3][\null]{\dfrac{\diff^{#1}{#2}}{\diff{#3}^{#1}}}              
\newcommand{\dpartial}[3][\null]{\dfrac{\partial^{#1}{#2}}{\partial{#3}^{#1}}}      

\newcommand{\hgm}{\hphantom{-}}

\newcommand{\gordre}[1]{\mathcal{O}\mathopen{}\left(#1\right)}

\newcommand{\Jd}{J_{2}}

\newcommand{\trunc}[1]{\left[#1\right]}   

\newcommand{\cpoiss}[2]{\left\{ #1 ; #2 \right\}}  

\DeclarePairedDelimiterX{\fracc}[2]{\lbrack}{\rbrack}{\dfrac{#1}{#2}}
\DeclarePairedDelimiterX{\fracp}[2]{\lparen}{\rparen}{\dfrac{#1}{#2}}
\DeclarePairedDelimiterX{\fracabs}[2]{\lvert}{\rvert}{\dfrac{#1}{#2}}

\DeclarePairedDelimiter{\abs}{\lvert}{\rvert}
\DeclarePairedDelimiter{\mean}{\langle}{\rangle}

\newcommand{\mca}[1]{\mathcal{#1}}
\newcommand{\mcat}[1]{\mca{\widetilde{#1}}}

\newcommand{\phy}{\varphi}

\renewcommand{\epsilon}{\text{\usefont{OML}{cmr}{m}{n}\symbol{15}}}

\newcommand{\anomelliptic}{\fontfamily{cmr}\selectfont \emph{w}}

\newcommand{\tagarray}{%
  \mbox{}\refstepcounter{equation}%
  $(\theequation)$%
}

\begin{document}

\delimitershortfall=-0.5pt
\delimiterfactor=901

\title{Towards an analytical theory of the third-body problem for highly elliptical orbits}

\author[1]{G.~Lion\thanks{Guillaume.Lion@oca.eu}}
\author[1]{G.~Métris\thanks{Gilles.Metris@oca.eu}}
\author[2]{F.~Deleflie\thanks{Florent.Deleflie@imcce.fr}}
\affil[1]{Géoazur, 
	Université de Nice Sophia-Antipolis, 
	CNRS (UMR 7329), 
	Observatoire de la Côte d’Azur,
	250 rue Albert Einstein, 
	Sophia Antipolis 06560 Valbonne, 
	France}
\affil[2]{IMCCE/GRGS, Observatoire de Paris, 
	CNRS (UMR 8028), 77 Avenue Denfert Rochereau 75014 Paris, France}

\renewcommand\Authands{ and }

\date{}

\maketitle

\begin{abstract}
  When dealing with satellites orbiting a central body on a highly elliptical orbit, it is necessary to consider the effect of gravitational perturbations due to external bodies. 
  Indeed, these perturbations can become very important as soon as the altitude of the satellite becomes high, which is the case around the apocentre of this type of orbit. 
  For several reasons, the traditional tools of celestial mechanics are not well adapted to the particular dynamic of highly elliptical orbits. 
  On the one hand, analytical solutions are quite generally expanded into power series of the eccentricity and therefore limited to quasi-circular orbits \cite{Giacaglia_1974aa, Kaula_1962aa}. 
  On the other hand, the time-dependency  due to the motion of the third-body is often neglected.
  We propose several tools to overcome these limitations. Firstly, we have expanded the disturbing function into a finite polynomial using Fourier expansions of elliptic motion functions in multiple of the satellite's eccentric anomaly (instead of the mean anomaly) and involving Hansen-like coefficients. 
  Next, we show how to perform a normalization of the expanded Hamiltonian by means of a time-dependent Lie transformation which aims to eliminate periodic terms. 
  The difficulty lies in the fact that the generator of the transformation must be computed by solving a partial differential equation involving variables which are linear with time and the eccentric anomaly which is not time linear. We propose to solve this equation by means of an iterative process.
\end{abstract}



\section{Introduction}

The problem of the expansion of the third body disturbing function has been extensively studied since a long time.

In 1959, Kozai \cite{Kozai_1959aa} developed a truncated theory limited to the second harmonic showing that the effects of the lunisolar perturbations may affect significantly the motion of artificial satellites. 
Later,  Musen et al. \cite{Musen_1961aa} took into account the third harmonic.
\\
Kaula \cite{Kaula_1961aa, Kaula_1966aa} introduced the inclination and eccentricity special functions, fundamental for the analysis of the perturbations of a satellite orbit. This enabled him to give in \cite{Kaula_1962aa} the first general expression of the third-body disturbing function using equatorial elements for the satellite and the disturbing body; 
the function is expanded using Fourier series in terms of the mean anomaly and the so-called Hansen coefficients depending on the eccentricity~$e$ in order to obtain perturbation fully expressed in orbital elements. 
\\
It was noticed by Kozai \cite{Kozai_1966aa} that, concerning the Moon, it is more suitable to parametrize its motion in ecliptic elements rather than in equatorial elements. Indeed, in this frame, the inclination of the Moon is roughly constant and the longitude of its right ascending node can be considered as linear with respect to time.
In light of this observation, Giacaglia and Bur{\v s}a~\cite{Giacaglia_1974aa}--\cite{Giacaglia_1980aa} established the disturbing function of an Earth's satellite due to the Moon attraction, using the ecliptic elements for the Moon and the equatorial elements for the satellite. However, by comparing their expressions with respect to the representation of the disturbing function in Cartesian coordinates, which is exact, we have noticed they are not correct.
Although Lane \cite{Lane_1989aa} highlighted some algebraic errors in \cite{Giacaglia_1974aa}, its development remains incorrect.
%
%
%
%
The main limitation of these papers is that they suppose truncations from a certain order in eccentricity. Generally, the truncation is not explicit because there is no explicit expansion in power of the eccentricity; but in practice Fourier series of the mean anomaly which converge slowly must be truncated and this relies mainly on the d'Alembert rule which guarantees an accelerated convergence as long as the eccentricity is small. Since this is indeed the case of numerous natural bodies or artificial satellites, these formulations are well suited in many situations. 
\\
However, there are also many examples of orbits of artificial satellites having very high eccentricities for which any truncation with respect to the eccentricity is prohibited.
Brumberg and Fukushima \cite{Brumberg_1994aa} investigated this situation.
They show that the series in multiples of the elliptic anomaly $\anomelliptic$, first introduced by Nacozy \cite{Nacozy_1977aa} and studied later by Janin and Bond \cite{Janin_1980aa}; Bond and Broucke \cite{Bond_1980aa}, converge faster than the series in multiples of any classical anomaly in many cases. These was confirmed by Klioner et al. \cite{Klioner_1997aa}. 
Unfortunately, the introduction of the elliptic anomaly increases seriously the complexity, involving in particular elliptical functions, e.g., see Dixon~\cite{Dixon_1894aa}.  
In the same paper, they give the expressions of the Fourier coefficients $Y_{s}^{n,m}$ and $Z_{s}^{n,m}$ in terms of hypergeometric functions, coming from the Fourier series expansion of the elliptic motion functions
in terms of the true anomaly and of the eccentric anomaly, respectively. 
More discussions and examples can be found in \cite{Brumberg_1999aa}.
\\
For completeness let us mention other researches developed by Da Silva Fernandes \cite{Da-Silva-Fernandes_1996aa} using a semi-analytical approach based on the expansions in powers of $(e-e_{0})$ in the neighborhood of a fixed value $e_{0}$. 
Note that this trick increases the radius of convergence of the Fourier coefficients in power series of the eccentricity, but does not improve the speed of convergence of the Fourier series.

The aim of the first part of the paper is to propose a new expression of the disturbing function which is in closed form with respect to the satellite eccentricity and still permits to construct an analytical theory of the motion. We will show that the use of the eccentric anomaly instead of the mean anomaly as fast angular variable fulfills this requirement. 

On the other hand, the expansion must be supple enough to define a trade-off between accuracy and complexity for each situation. 
To this end, the use of special functions is well suited: the expansions are compact and easy to manipulate and the extension of the theory is chosen for each case by fixing the limits of the summations. The complexity is relegated in the special functions, knowing that efficient algorithms exist to compute these functions. In short, we shall develop an expression of the disturbing function mixing mainly the compactness of the formulation of  Giacaglia and Burša \cite{Giacaglia_1980aa} in exponential form and the convergence of Fourier series of the eccentric anomaly. 

Even if some formulas written in this paper are already published in previous article, we think that for clarity and consistency, it is useful to present the developments from the beginning.

Besides the question of large eccentricities, the other issue concerning the third body perturbation is that it is explicit time-dependency. 
This should be taken into account  when constructing an analytical solution, in particular by means of canonical transformations. 
To do this, the key point is to start from a disturbing function using angular variables which are time linear. 
This is precisely the motivation to use ecliptic elements instead of equatorial elements for the Moon perturbation, as explained above. 
In this situation, the PDE (Partial Differential Equation) that we have to solve to construct an analytical theory are of the form:%
\begin{equation*}
  \sum_{i=1}^n \omega_i \dpartial{\mca{W}}{\alpha_i} = {}
  \mca{A} \cos \left( \sum_{i=1}^n   k_i \alpha_i \right) \;,
\end{equation*}
with the obvious solution
\begin{equation*}
  \mca{W} = {} 
  \frac{\mca{A}}{\mathchar"1358\limits_{i=1}^n k_i \omega_i} \sin \left( \sum_{i=1}^n  k_i \alpha_i\right) \;.
\end{equation*}
Unfortunately, this nice mechanics is broken as soon as the fast variable of the satellite motion is no longer the mean anomaly $M$, but instead is the eccentric anomaly $E$. The equation to solve in this case looks like%
\begin{equation*}
  \omega_1\dpartial{\mca{W}}{M} + \sum_{i=2}^n \omega_i \dpartial{\mca{W}}{\alpha_i} = {}
  \mca{A} \cos \left( k_1 E + \sum_{i=2}^n   k_i \alpha_i\right) \;,
\end{equation*}
the solution of which is not trivial. In the second part of the paper, we propose a way to deal with this difficulty.

The paper is organized as follows: in Section \ref{sec-Model}, we present the hierarchy of the perturbations acting on a satellite and we define the Hamiltonian used in our analytical theory of the third-body problem. 
Sections \ref{sec-EMDF}-\ref{sec-ESDF} are devoted to the development of the Moon and Sun disturbing functions fully expressed in orbital elements, satisfying our objectives. 
Particularly, we recall the expressions in spherical coordinates and in Hill-Whittaker variables: these are more or less well known results but they are expressed in a form suited for our use. 
In Section \ref{sec-APPS}, we show how our development can be used in a canonical Lie Transform, in order to eliminate all angular variables.


\section{Model} 
\label{sec-Model}

In an inertial geocentric reference frame, we study the perturbations on a satellite orbit due to the Earth's oblateness, the Earth's gravity field and the attraction of external bodies (e.g. Moon, Sun).
The satellite (mass $M$) and the third-body (mass $M'$)  are respectively located at a radial distance $r$ and $r'$ from the Earth's centre. 
We suppose that $r'/r>1$ and $M'/M \gg 1$. 
Here and subsequently, we will use unprimed variables for satellite and primed variables for the disturbing body.

\subsection{Hierarchy of the perturbations}
\label{sec-Analyse_pert}

\begin{figure}[!htb]
  \centering
  \includegraphics[clip=true, trim = 30 0 0 20, width=\linewidth]{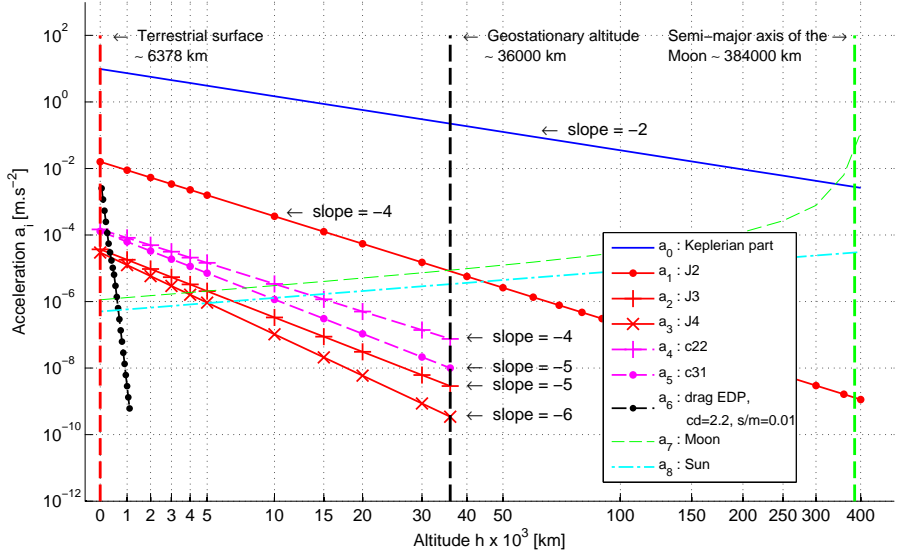}
  \caption{Order of magnitude of the acceleration experienced by a celestial body arising from gravitational and non-gravitational perturbations as function of the altitude from the mean terrestrial surface. The axes are in log scale.}
  \label{fig:acc}
\end{figure}

In Figure \ref{fig:acc}, we have plotted the evolution with respect to the distance, of the different parts of the gravitational acceleration:%
\begin{itemize}
\item the keplerian part;
\item the non-sphericity of the Earth: zonal harmonic terms: $\Jd, J_{3}, J_{4}$ and tesseral harmonic terms: $C_{21}, C_{31}$;
\item lunisolar perturbations.
\end{itemize}
The evolution of the atmospheric drag is also plotted using an exponential density profile~(EDP). 

We have the following hierarchy depending on the altitude:
\begin{itemize}
\item \textbf{Low altitude}: the dominant perturbation is obviously $\Jd$, then the atmospheric drag;
\item \textbf{Medium altitude}: the dominant perturbation is always $\Jd$, followed by the lunisolar perturbations;
\item \textbf{High altitude}: in this region, the lunisolar perturbations can reach or exceed the same order of magnitude as $\Jd$.
\end{itemize}

Since we are interested in studying the dynamics for highly elliptical orbits, and in order to simplify the problem, only the oblateness term and the third-body effects are considered. 
Indeed, a satellite on this kind of orbit spends most of the time at a high altitude. In terms of semi-major axis $a$ and eccentricity $e$, we must have: $a \ge R_{\varoplus}/(1-e)$, with $R_{\varoplus}$ the equatorial radius of the Earth.

\subsection{The Hamiltonian formalism}
\label{sec-Hamiltonien}

We introduce the osculating orbital elements: $a$, $e$, $I$ the inclination, $\Omega$ the longitude of the ascending node, $\omega$ the argument of periapsis and $M$ the mean anomaly; and the classical Delaunay canonical variables:%
\label{eq:eq_Delaunay}
\begin{alignat}{4}
  & L 
  && = {} \sqrt{\mu\,a} 	
  && {} \quad ; \quad {}  l 
  && = {} M \;,
  \nonumber
  \\
  & G 
  && = {} \eta L
  && {} \quad ; \quad {} g 
  && = {} \omega \;,
  \\
  & H
  && = {} G \cos I 
  && {} \quad ; \quad {} h 
  && = {} \Omega \;.
  \nonumber
\end{alignat}
where $\mu = \mca{G} M_{\varoplus} $ is the geocentric gravitational constant, $\mca{G}$ is the gravitational constant, $M_{\varoplus}$ the mass of the Earth and $\eta = \sqrt{1-e^{2}}$. For compactness, we will note $y=(l,g,h)$ and $Y=(L,G,H)$.

The Hamiltonian $\mca{H}$ of the system can be expressed as a sum of two terms:%
\begin{equation}
  \label{eq:H_J2_mod}
  \mca{H} = {} 
  \mca{H}_{0}+ \mca{H}_{1}\;.
\end{equation}
We have chosen to put in the main part $\mca{H}_{0}$ of the Hamiltonian, the keplerian part $\mca{H}_{kep}$ and the secular variations of the oblateness term in $\Jd$, noted $\mca{H}_{\Jd, sec}$:%
\begin{subequations}
  \begin{alignat}{2}
    \label{eq:H0}
    \mca{H}_{0} 
    & = {} \mca{H}_{0} (\_,Y) = {} \mca{H}_{kep}(\_,\_ ,\_ ,L,\_,\_) + \mca{H}_{\Jd}(\_,\_,\_,L,G,H) \;,
    \\
    \label{eq:Hkep}
    \mca{H}_{kep} 
    & =  {-} \frac{\mu^2}{2\,L^2} \;,
    \\
    \label{eq:HJ2sec}
    \mca{H}_{\Jd, sec} 
    & = {} \frac{n_{0}^{2}}{4} \Jd R_{\varoplus}^2 \fracp*{L}{G}^3 
    \left[ 1 - 3 \fracp*{H}{G}^2
    \right] \;,
  \end{alignat}
\end{subequations}
where $n_{0}$ is the mean motion defined by,
\begin{equation}
  \label{eq:n0}
  n_{0} 
  = {} \frac{\mu^2}{L^3} \:.
\end{equation}
The perturbing part $\mca{H}_{1}$, of order 1, contains in principle the short periodic variations in~$\Jd$, noted $\mca{H}_{\Jd, per}$, and the third-body perturbations $\mca{H}_{3c}$ (Moon, Sun), which can reach the same order of magnitude than~$\Jd$ (see Figure~\ref{fig:acc}):%
\begin{equation}
  \label{eq:H1}
  \mcat{H}_{1} 
  = {} \mca{H}_{\Jd, per} + \mca{H}_{3c} \;.
\end{equation}
In fact, periodic perturbations due to $\Jd$ are not considered in this paper (this point will be discussed in the conclusion), and we retain only the part:%
\begin{equation}
  \label{eq:def_H1}
  \mca{H}_{1}
  = {} \mca{H}_{1} (y,y',Y,Y')
  = {} \mca{H}_{3c} \;.
\end{equation}
%
Note that, this perturbed Hamiltonian is explicitly time dependent through the position of the disturbing body.
The Hamilton equations are written as follows:
\begin{alignat}{4}
  \label{eq:sys_eq_Delaunay}
  & \derive{L}{t} 
  && {} = {-}\dpartial{\mca{H}}{l} 	
  && {} \quad ; \quad {} \derive{l}{t} 
  && {} = {} \dpartial{\mca{H}}{L} \;,
  \nonumber
  \\
  & \derive{G}{t} 
  && {} = {-}\dpartial{\mca{H}}{g} 	
  && {} \quad ; \quad {} \derive{g}{t} 
  && {} = {} \dpartial{\mca{H}}{G} \;,
  \\
  & \derive{H}{t} 
  && {} = {-}\dpartial{\mca{H}}{h} 	
  && {} \quad ; \quad {} \derive{h}{t} 
  && {} = \dpartial{\mca{H}}{H} \;.
  \nonumber
\end{alignat}


\section{Moon disturbing function} \label{sec-EMDF}

\subsection{Spherical harmonics and transformation under rotations} \label{sec-SH_rot}

Let $\mathscr{S}$ (O,x,y,z) be a reference frame and $P$ a point of spherical coordinates: $r$ the radius, $\theta$ the co-latitude and $\phy$ the longitude.
For any couple of integers ($n$, $m$) with $n \geq 0$ and $-n \leq m \leq n$, we define the normalized complex spherical harmonics $\overline{Y}_{n,m}(\theta,\phy)$ by~
\begin{subequations}
  \label{eq:HS_conv_Ferrers}
  \allowdisplaybreaks[4] 
  \begin{align}
    & \overline{Y}_{n,m}(\theta, \phy) = {} N_{n,m} Y_{n,m}(\theta, \phy) \;, \\
    & Y_{n,m}(\theta, \phy) = {} P_{n,m}(\sin \theta) \exp \ci m \phy \;, \\
    & N_{n,m} = {} \sqrt{ (2n+1) \frac{(n-m)!}{(n+m)!} } \;,         
  \end{align}
\end{subequations}
where the $P_{n,m}(\sin \theta) $ are the associated Legendre polynomials of degree $n$ and order $m$ (see e.g. Abramowitz M. and Stegun I.~A. \cite{Abramowitz_1972aa}), $N_{n,m}$ is the much more common Ferrers normalization factor \cite{Ferrers_1877aa} (geodesy convention), and $\ci = \sqrt{-1}$. 

Note that for $m \leq 0$, we have the property,
\begin{equation}
  \label{eq:Pnm_m_neg}
  P_{n,-m}(\sin \theta) = (-1)^{m} \frac{(n-m)!}{(n+m)!} P_{n,m}(\sin \theta) ,
\end{equation}
which then also implies that
\begin{equation}
  \label{eq:Ynm_m_neg}
  Y_{n,-m}(\theta, \phy) = {} (-1)^{m} \frac{(n-m)!}{(n+m)!}\, Y_{n,m}^{\ast}(\theta, \phy) \;,
\end{equation}
where the asterisk "*" denotes complex conjugation.

Consider now a new frame $\mathscr{S'}$ $(O, x',y',z')$ and  $( \theta', \phy')$ the new angular spherical coordinates of $P$ in this frame. 
If the transformation between the two frames is a 3-1-3 Euler angle sequence $( \alpha-\frac{\pi}{2}, \beta, \gamma+\frac{\pi}{2})$, the spherical harmonics transform under rotations as (see Sneeuw \cite{Sneeuw_1992aa}):%
\begin{equation}
  \label{eq:1}
  Y_{n,m}(\theta, \phy) =
  \sum_{k = -n}^{n} 
  D_{n,m,k} \left( \alpha-\frac{\pi}{2},\beta,\gamma+\frac{\pi}{2} \right) 
  Y_{n,k}(\theta', \phy') \;.
\end{equation}
The coefficients of the transformation are defined by 
\begin{subequations}
  \begin{align}
    \label{eq:EARS}
    D_{n,m,k} \left( \alpha-\frac{\pi}{2}, \beta,\gamma+\frac{\pi}{2} \right) 
    & = {} \exp\ci m \left( \alpha-\frac{\pi}{2} \right) d_{n,m,k} (\beta) \exp\ci k \left( \gamma+\frac{\pi}{2} \right) \;,
    \\
    & = \ci^{k-m} \, d_{n,m,k} (\beta) \, \exp \ci \left( k\gamma + m\alpha \right) \;,
  \end{align}
\end{subequations}
where $d_{n,m,k} (\beta)$ are the real coefficients given explicitly by the Wigner formula\footnote{The original Wigner formula \cite{Wigner_1959aa} applies to a 3-2-3 sequence, while in our context, we use a 3-1-3 sequence; a 3-1-3 rotation sequence is equivalent to a 3-2-3 sequence providing  we subtract $\pi/2$ to the first rotation and we add $\pi/2$ to the third one.}:
\begin{subequations}
  \label{eq:Wigner_d}
  \begin{align}
    \label{eq:Wigner_U2d}
    d_{n,m,k} (\beta) 
    & =  (-1)^{k-m} \frac{(n-k)!}{(n-m)!} U_{n,m,k} (\beta) \;,
    \\
    \label{eq:Wigner_U}
    \begin{split}
      U_{n,m,k} (\beta) 
      & = (-1)^{n-k} \sum_{r}^{}  (-1)^{r} \binom{n-m}{r} \binom{n+m}{m+k+r} 
      \\ 
      & \hspace{2em} \times
      \cos^{a} \fracp*{\beta}{2} \, \sin^{2n-a} \fracp*{\beta}{2} \;,
    \end{split}
  \end{align}
\end{subequations}
%
with $a = 2r+m+k$ where $r$ ranges from ${ \max(0,-k-m)}$ to $ \min(n-k, n-m)$. 
Here, the $U$-functions are the same used in Giacaglia \cite{Giacaglia_1974aa} and Lane \cite{Lane_1989aa}.


\subsection{Expansion in spherical coordinates}\label{sec-Exp_Moon_sph}

Introducing the equatorial coordinates $(\alpha,\delta)$ and using the addition formula of Legendre polynomials fully denormalized (see e.g. \cite{Gradshteuin_2007aa}, Eq. (8.814)), the disturbing function $\mca{R}_{3c}$ for an external body can be written
\begin{equation}
  \label{eq:R_YYb}
  \mca{R}_{3c} 
  = {} \frac{\mu'}{r'} 
  \sum_{n \geq 2}^{} \sum_{m = -n}^{n} 
  \frac{(n-m)!}{(n+m)!} \fracp*{r}{r'}^{n}
  Y_{n,m}(\delta, \alpha) Y^{\ast}_{n,m}(\delta', \alpha') \;,
\end{equation}
where $\mu'=\mca{G} M'$.

Note that $\mca{R}_{3c}$ is the opposite of the potential energy and thus $\mca{H}_{3c} = - \mca{R}_{3c}$.


\subsection{Expansion in Hill-Whittaker variables}\label{sec-Exp_Moon_HW}

In order to construct an analytical theory, it is suitable to express the disturbing potential as function of the osculating orbital elements or equivalent variables.

Giacaglia \cite{Giacaglia_1974aa} established the first general expression of the lunar disturbing function in terms of ecliptic elements.
As he wrote later, his approach to the problem is messy and obscure. 
Having noticed some error calculations leading to relation~15, Lane \cite{Lane_1989aa} proposed a new expression, but still not correct. 
Calculations in his paper are not trivial and require a little "algorithm" to restore signs of trigonometric terms due to the fact that Lane converts powers of the complex number $\ci^{k}$ by powers of $(-1)^{\trunc{k/2}}$ (which is valid only for non-negative integer values of~$k$).

In order to obtain a valid trigonometric formulation of the lunar disturbing body, we propose to use and convert an exponential formulation. Such a formulation is given in Giacaglia and Burša \cite{Giacaglia_1980aa}, Eq. (35), but the normalization convention and some functions do not correspond to those used widely in Celestial Mechanics. 
For better understanding and more clarity, we have chosen to present the main steps of the demonstration with the standard convention. 

An other proof is given in Appendix \ref{An-Method_2} where we have taken the approach used by Giacaglia and Burša \cite{Giacaglia_1980aa}.

\subsubsection{\textbf{Exponential formulation}.}

In a first step, we express \eqref{eq:R_YYb} as function of the Hill-Whittaker variables $r$, $\dot r$, $\theta=\omega+\nu$, $G=\sqrt{\mu a (1-e^2)}$, $\Omega$ and $H=G\cos I$ with $\nu$ the true anomaly (see Whittaker \cite{Whittaker_1904aa} and Deprit \cite{Deprit_1980aa}).
For conciseness we will still use the inclination $I$ instead of $H/G=\cos I$.

We begin by expressing the complex spherical harmonics of the satellite $Y_{n,m}(\delta, \alpha)$ in terms of osculating orbital elements. 
Such a transformation is equivalent to a rotation of the equatorial frame into the orbital frame. 
This can be written using the generalized inclination functions $F_{n,m}^k(I)$ (see Gooding and Wagner \cite{Gooding_2008aa}):~
\begin{subequations}
  \allowdisplaybreaks[4] 
  \label{eq:Gooding_Fnmk}
  \begin{flalign}
    & Y_{n,m}(\delta, \alpha) = {} 
    \ci^{n-m} \sum_{k=-n}^{n} F_{n,m}^{k} (I) \exp\ci\Psi_{m,k} \;,
    \label{eq:Ynm_Sat_eo2eq_Fnmk}
    \\
    & F_{n,m}^{k} (I) = (-1)^{\frac{n-k}{2}} d_{n,m,k} (I) P_{n,k}(0) \;,
    \\
    & P_{n,k} (0) =
    \label{eq:def_Pnm_{0}}
    \begin{cases}
      (-1)^{(n-k)/2} \dfrac{(n+k)!}{2^n [(n-k)/2]! [(n+k)/2]!} & ,  \text{ for ($n-k$) even}
      \\
      0  & ,  \text{ for ($n-k$) odd}
    \end{cases}
  \end{flalign}
\end{subequations}
with
\begin{equation}
  \label{eq:3}
  \Psi_{m,k} = k \theta + m \Omega \;,
\end{equation}
and $d_{n,m,k}(I)$ are defined in \eqref{eq:Wigner_d}.

Replacing $k$ by $n - 2p$ leads to
\begin{equation}
  \label{eq:Ynm_Sat_eo2eq_Fnmp}
  Y_{n,m}(\delta, \alpha) = {} 
  \ci^{n-m} \sum_{p=0}^{n} F_{n,m,p} (I) \exp\ci\Psi_{n,m,p} \;,
\end{equation}
with 
\begin{equation}
  \label{eq:6}
  \Psi_{n,m,p} = (n-2p) \theta + m \Omega \;,
\end{equation}
and $F_{n,m,p} (I) = F_{n,m}^{n-2p} (I)$. 
These latter are related to the Kaula's inclination functions \linebreak[3]{\cite{Kaula_1961aa, Kaula_1966aa}} by the factor $(-1)^{\trunc{(n-m+1)/2}}$. We refer the reader to Gooding and Wagner \cite{Gooding_2008aa,Gooding_2010aa} for more details (background and numerical computation) on the $F$ and $d$-functions. 

We note that  Brumberg et al. \cite{Brumberg_1971aa} and Brumberg \cite{Brumberg_1995aa} propose a relation equivalent to the transformation \eqref{eq:Gooding_Fnmk}.

To parametrize the motion of the Moon using linear angular variables with respect to the time, it is advised to use orbital elements referred to the ecliptic plane (see Kozai \cite{Kozai_1966aa}).
Introducing the ecliptic longitude $\lambda'$ and the ecliptic latitude $\beta'$, the spherical harmonics $Y_{n,m}(\delta', \alpha')$ transform under rotation as (see Eq. \eqref{eq:1}):
\begin{equation}
  \label{eq:Ynm_Moon_ec2eq}
  Y_{n,m}(\delta', \alpha') = {} 
  \sum_{m'=-n}^{n}  
  \ci^{m'-m} d_{n,m,m'}^{} (\epsilon) Y_{n,m'}(\beta', \lambda') \;,
\end{equation}
where $\epsilon$ is the obliquity. 

Spherical harmonics $Y_{n,m'}(\beta', \lambda')$ are then expressed into lunar orbital frame thanks again to the relation \eqref{eq:1} using a 3-1-3 Euler  sequence involving the coefficients $D_{n,m,m'} \left( \Omega'-\frac{\pi}{2},I',\theta'+\frac{\pi}{2} \right)$. 
By construction, the longitude and latitude of the moon in the new frame are zero, and we obtain
\begin{equation}
  \label{eq:Ynm_Moon_eo2ec_dnmk}
  Y_{n,m'}(\beta', \lambda') =\sum_{r = -n}^{n} 
  \ci^{r-m'} d_{n,m',r} (I') Y_{n,r}(0,0) \exp{\ci( m' \Omega' + r \theta')} \;.
\end{equation}
Knowing that $Y_{n,r}(0,0) = P_{n,r}(0)$ (see Eq. \eqref{eq:def_Pnm_{0}}), we replace $r$ by $n-2p'$, so that
\begin{equation}
  Y_{n,m'}(\beta', \lambda') = \ci^{n-m'}
  \sum_{p' = 0}^{n} F_{n,m',p'}(I') \exp\ci \Psi'_{n,m',p'} \;,
  \label{eq:Ynm_Moon_eo2ec_Fnmp}
\end{equation}
and
\begin{equation}
  \label{eq:Ynm_Moon_eo2eq}
  Y_{n,m}(\delta', \alpha') = {}
  \ci^{n-m}
  \sum_{m'=-n}^{n}  
  \sum_{p' = 0}^{n}  
  d_{n,m,m'}^{} (\epsilon) 
  F_{n,m',p'}(I') \exp\ci \Psi'_{n,m',p'} \;,
\end{equation}
with 
\begin{equation}
  \label{eq:8}
  \Psi'_{n,m',p'} 
  = (n-2p') \theta' + m' \Omega' \;.
\end{equation}

Substituting \eqref{eq:Ynm_Sat_eo2eq_Fnmp} and \eqref{eq:Ynm_Moon_eo2eq} into \eqref{eq:R_YYb}, and using the relation \eqref{eq:Wigner_U2d} to convert the $d$-functions into $U$-functions, the disturbing function takes the form
\begin{align}
  \label{eq:R_moon_HK_exp_Fnmp}
  \begin{split}
    \mca{R}_{\leftmoon} & = {}
    \frac{\mu'}{r'}  \sum_{n \geq 2}^{}
    \sum_{m = -n}^{n} \,
    \sum_{m' = -n}^{n} \,
    \sum_{p = 0}^{n} \,
    \sum_{p' = 0}^{n} 
    \fracp*{r}{r'}^{n} 
    (-1)^{m-m'} \frac{(n-m')!}{(n+m)!} 
    \\
    {} & \hspace{2em} \times
    F_{n,m,p}(I) F_{n,m',p'}(I') U_{n,m,m'}(\epsilon)
    \exp\ci \left( \Psi_{n,m,p} - \Psi'_{n,m',p'} \right) \;.
  \end{split}
\end{align}

In Table \ref{tab_comparaison}, we give the relations between the functions used by  Giacaglia and Burša~\cite{Giacaglia_1980aa} and the classical functions used in this paper yielding to the lunar disturbing function~\eqref{eq:R_moon_HK_exp_Fnmp}.

{
  \renewcommand{\arraystretch}{1.25}

  \begin{table}[h!]


    \begin{flushleft}
      \hfill
      \begin{minipage}[t]{0.80\textwidth}

        \centering
        \begin{tabular}{%
            *{1}{L{9em}} 
            *{1}{K{1em}} 
            *{1}{R{14em}}}    
          
          \toprule
          
          \text{Giacaglia and Burša} & &  \text{Our relations}  \\

          \cmidrule(l ){1-1}
          \cmidrule( r){3-3}

          F_{n,m,p}(I) & = {} & \ci^{n-m} F_{n,m,p}(I)  \\
          D_{n,-m,-m'}(0,\epsilon,0) & = {} & D_{n,m,m'}(-\pi/2,\epsilon,\pi/2) \\
          F'_{n,m',p'}(I') & = {} & (-1)^{n} \ci^{m'-n} F_{n,-m',n-p'}(I') \\
          & = {} & (-1)^{m'} \ci^{m'-n} \frac{(n-m')!}{(n+m')!} F_{n,m',p'}(I') \\
          K_{n,m,m'}(\epsilon) & = {} & (-1)^{m} \ci^{m-m'} d_{n,m,m'}(\epsilon) \\
          
          \cmidrule(l ){1-1}
          \cmidrule( r){3-3}
          
          \mca{R}_{n,m,m',p,p'}^{\prime} &  \equiv{} & \frac{(n+m)!}{(n-m)!} \mca{R}_{n,m,m',p,p'}  \\

          \bottomrule
          
        \end{tabular}

        \caption{Comparison between functions used in Giacaglia and Burša \cite{Giacaglia_1980aa} and those of our paper yielding to the disturbing function of the Moon \eqref{eq:R_moon_HK_exp_Fnmp}.}
        \label{tab_comparaison}

      \end{minipage}
      \hfill
      \begin{minipage}[t]{0.11\textwidth}
        \begin{tabular}{Cr}    
          
          \rule[-11pt]{0pt}{0pt}
          {} & \\

          \vphantom{\ci^{n}} & \tagarray\label{eq:comp_Ta}  \\
          \vphantom{D_{m'}} & \tagarray\label{eq:comp_Tb} \\
          \vphantom{\ci^{m'}} & \\
          \vphantom{\frac{(n)!}{(n)!}} & \tagarray\label{eq:comp_Tc}  \\
          \rule[-11pt]{0pt}{0pt}
          \vphantom{\ci^{m'}} & \tagarray\label{eq:comp_Td} \\
          \vphantom{\frac{(n)!}{(n)!}} & \tagarray\label{eq:comp_Te} \\


        \end{tabular}
      \end{minipage}
    \end{flushleft}

  \end{table}
}

The main difference between our formulation of $\mca{R}_{\leftmoon}$ and those of \cite{Giacaglia_1980aa} comes from the convention used to define the addition formula for Legendre polynomials (see the relation~\eqref{eq:comp_Te} in Table \ref{tab_comparaison}). In our case, we use a Ferrers normalization \eqref{eq:HS_conv_Ferrers} while Giacaglia seems to use Schmidt normalized.

\subsubsection{\textbf{Trigonometric formulation}.}

We propose here a fast and simple method to convert the disturbing function into trigonometric form from the relation \eqref{eq:R_moon_HK_exp_Fnmp}.

Let 
\begin{subequations}
  \begin{align}
    \mca{R} & = {} \frac{\mu'}{r'} \sum_{n \geq 2}^{} \, \sum_{p = 0}^{n} \, \sum_{p' = 0}^{n} \fracp*{r}{r'}^{n} \mca{R}_{n,p,p'}
    \;,
    \\
    \begin{split}
      \mca{R}_{n,p,p'} & = {} \sum_{m = -n}^{n} \, \sum_{m' = -n}^{n} (-1)^{m-m'} \frac{(n-m')!}{(n+m)!}
      \\
      {} & {} \hspace{2em} \times F_{n,m,p}(I) F_{n,m',p'}(I') U_{n,m,m'}(\epsilon) \exp\ci \left( \Psi_{n,m,p} -
        \Psi'_{n,m',p'} \right) \;.
    \end{split}
  \end{align}
\end{subequations}
At first, we split the sum over $m$ and $m'$ into four parts such that $m$ and $m'$ are positive or null,%
\begingroup\footnotesize
\setlength\abovedisplayskip{0pt}
\begin{equation}
  \label{eq:R_Moon_etape_2}
  \begin{split}
    \mca{R}_{n,p,p'} & = 
    {} \sum_{m = 0}^{n} \, \sum_{m' = 0}^{n} \,
    \frac{ \Delta_{0}^{m,m'} }{2} 
    (-1)^{m-m'}
    \\ {} & {} \hspace{2em} \times
    \left[
      \frac{(n-m')!}{(n+m)!} F_{n,m,p}(I) F_{n,m',p'}(I') U_{n,m,m'}(\epsilon)
      \exp\ci \left( \Psi_{n,m,p} - \Psi'_{n,m',p'} \right)
    \right. \\ {} & {} \left.
      \hspace{3em} + {}
      \frac{(n+m')!}{(n+m)!} F_{n,m,p}(I) F_{n,-m',p'}(I') U_{n,m,-m'}(\epsilon)
      \exp\ci \left( \Psi_{n,m,p} - \Psi'_{n,-m',p'} \right)
    \right. \\ {} & {} \left.
      \hspace{3em}+ {}
      \frac{(n-m')!}{(n-m)!} F_{n,-m,p}(I) F_{n,m',p'}(I') U_{n,-m,m'}(\epsilon)
      \exp\ci \left( \Psi_{n,-m,p} - \Psi'_{n,m',p'} \right)
    \right. \\ {} & {} \left.
      \hspace{3em}+ {}
      \frac{(n+m')!}{(n-m)!} F_{n,-m,p}(I) F_{n,-m',p'}(I') U_{n,-m,-m'}(\epsilon)
      \exp\ci \left( \Psi_{n,-m,p} - \Psi'_{n,-m',p'} \right) 
    \right] \;,
  \end{split}
\end{equation}
\endgroup
where
\begin{equation}
  \label{eq:Delta}
  \Delta_{0}^{m,m'} = {} \frac{(2 - \delta_{0}^{m}) \, (2 - \delta_{0}^{m'})}{2} \;,
\end{equation}
in which, $\delta_{j}^{k}$ is the Kronecker delta.  

We may find in Sneeuw \cite{Sneeuw_1992aa} some interesting symmetry properties about the inclination functions and the coefficients $d_{n,m,m'}$ for negative values of $m$ or $m'$:
\begin{subequations}
  \label{eq:prop_fonc_spe}
  \begin{align}
    U_{n,-m,-m'} & = {} (-1)^{m'-m} U_{n,m,m'} \;,
    \\
    F_{n,-m,n-p} & = {} (-1)^{n-m} \frac{(n-m)!}{(n+m)!} F_{n,m,p} \;.
  \end{align}
\end{subequations}
Additionally, we have the property,
\begin{equation}
  \label{eq:prop_Psi}
  \Psi_{n,-m,n-p} = {-} \Psi_{n,m,p} \;.
\end{equation}
Changing $p$ by $n-p$ in the third and fourth term of \eqref{eq:R_Moon_etape_2}, then $p'$ by $n-p'$ in the second and fourth term, we substitute each function for which the second index has a negative sign by their equivalent expression given above. After some rearrangement, we find%
\begingroup\footnotesize
\setlength\abovedisplayskip{0pt}
\begin{equation}
  \label{eq:R_Moon_etape_6}
  \begin{split}
    \mca{R}_{n,p,p'} & = {} 
    \sum_{m = 0}^{n} \sum_{m' = 0}^{n} \,
    \frac{ \Delta_{0}^{m,m'} }{2} 
    \frac{(n-m')!}{(n+m)!} F_{n,m,p}(I) F_{n,m',p'}(I')
    \\ {} & {} \hspace{2em} \times 
    \left[
      (-1)^{m-m'} U_{n,m,m'}(\epsilon)
      \left( \exp\ci\Theta_{n,m,m',p,p'}^{-} + \exp \left(-\ci\Theta_{n,m,m',p,p'}^{-} \right) \right)
    \right. \\ {} & {} \left. \hspace{3em}
      + (-1)^{n-m}  U_{n,m,-m'}(\epsilon)
      \left( \exp\ci\Theta_{n,m,m',p,p'}^{+} + \exp \left(-\ci\Theta_{n,m,m',p,p'}^{+} \right) \right)
    \right] \;,
  \end{split}
\end{equation}
\endgroup
where $\Theta_{n,m,m',p,p'}^{\pm} = {} \Psi_{n,m,p} \pm \Psi'_{n,m',p'}$.

Finally, converting the exponential terms to trigonometric terms, the lunar disturbing function thus becomes%
\begingroup\small
\setlength\abovedisplayskip{0pt}
\begin{align}
  \label{eq:R_moon_HK_trig_Fnmp}
  \begin{split}
    \mca{R}_{\leftmoon} & = {}
    \frac{\mu'}{r'}  \sum_{n \geq 2}^{}
    \sum_{m = 0}^{n} \,
    \sum_{m' = 0}^{n} \,
    \sum_{p = 0}^{n} \,
    \sum_{p' = 0}^{n}
    \Delta_{0}^{m,m'} (-1)^{m-m'}
    \frac{(n-m')!}{(n+m)!} 
    \fracp*{r}{r'}^{n} 
    F_{n,m,p}(I) F_{n,m',p'}(I') 
    \\ 
    {} & {} \hspace{2em} \times
    \left[ U_{n,m,m'}(\epsilon) \cos \Theta_{n,m,m',p,p'}^{-}
      + (-1)^{n-m'} U_{n,m,-m'}(\epsilon) \cos \Theta_{n,m,m',p,p'}^{+} \right] \;.
  \end{split}
\end{align}
\endgroup
As we can see, this expression involves only cosine terms contrary to the mistaken formulation given in Giacaglia and Lane.


\subsection{Expansion in osculating orbital elements} \label{sec-Exp_Moon_OE}

Let us introduce the elliptic motion functions
\begin{equation}
  \label{eq:_Phi_nk}
  \Phi_{n,k} = {} 
  \fracp*{r}{a}^{n} \exp\ci k\nu \;.
\end{equation}

Expressions \eqref{eq:R_moon_HK_exp_Fnmp} still depend on $r$, $r'$, $\nu$ and $\nu'$ (via $\theta$ and $\theta'$). 
To obtain a perturbation fully expressed in orbital elements, the classical way is to introduce expansions in Fourier series of the mean anomaly of the form:
\begin{equation}
  \Phi_{n,k} = \sum_{q=-\infty}^{+\infty} X_{q}^{n,k}(e) \exp\ci qM \;,
  \label{eq:dev_fourier_M}
\end{equation}
where $X_{q}^{n,k}(e)$ are the well known Hansen coefficients \cite{Hansen_1853aa}. 
In the general case, the series~\eqref{eq:dev_fourier_M} always converge as Fourier series but can converge rather slowly (see e.g. Klioner et al. \cite{Klioner_1997aa} or Brumberg and Brumberg \cite{ Brumberg_1999aa}). 
Only in the particular case where~$e$ is small,  the convergence is fast thanks to the d'Alembert property which ensures that~$e^{\abs{k-q}}$ can be factorized in $X_{q}^{n,k}(e)$. 
That is why the method is reasonably efficient for most of the natural bodies (in particular the Sun and the Moon) but can fail for satellites moving on orbits with high eccentricities. 
In this case, Fourier series of the eccentric anomaly $E$ (see Brumberg and Fukushima \cite{Brumberg_1994aa}) are much more efficient:%
\begin{equation}
  \Phi_{n,k} = \sum_{q=-\infty}^{+\infty} Z_{q}^{n,k}(e) \exp\ci q E \;,
  \label{eq:dev_fourier_E_1}
\end{equation}
In case where $0\leq |k| \leq n$, the coefficients $Z_{q}^{n,k}(e)$ can be expressed in closed form and  the sum over $q$ \eqref{eq:dev_fourier_E_1} is bounded by $\pm{}n$ (coefficients are null for $|q|>n$). The expression of the Fourier coefficients $Z_{q}^{n,k}(e)$ are given in Annexe \ref{An-Znms} only for this particular case. Other general expressions and numerical methods to compute them can be find in Klioner et al.~\cite{Klioner_1997aa}, Laskar~\cite{Laskar_2005aa}, Lion and Métris \cite{Lion_2013aa}.

Even if this kind of expansion does not allow to express the disturbing function strictly in orbital elements, the key point is that the required operations (derivation and integration with respect to the mean anomaly) can be easily performed thanks to the relation
\begin{equation}
  \label{eq:20}
  \mathrm{d}M =\frac{r}{a} \mathrm{d}E \;.
\end{equation}

Rewriting the ratio of the distance $r$ to $r'$ within \eqref{eq:R_moon_HK_exp_Fnmp} as
\begin{equation}
  \label{eq:ratio a_ap}
  \frac{1}{r'} \fracp*{r}{r'}^{n} = {} 
  \frac{1}{a'} \fracp*{a}{a'}^n 
  \fracp*{a}{r}  \fracp*{r}{a}^{n+1} \fracp*{a'}{r'}^{n+1}  \;,
\end{equation}
we get 
\begin{subequations}
  \allowdisplaybreaks[4] 
  \label{eq:R_moon_EO_exp_Fnmp}
  \begin{align}
    & \mca{R}_{\leftmoon} = {}
    \sum_{n \geq 2}^{} \,
    \sum_{m = -n}^{n} \,
    \sum_{m' = -n}^{n} \, 
    \sum_{p = 0}^{n} \,
    \sum_{p' = 0}^{n} \,
    \sum_{q = -n-1}^{n+1} \, 
    \sum_{q' = -\infty}^{+\infty} \, 
    \mca{R}_{n,m,m',p,p',q,q'} \;,
    \\
    & \mca{R}_{n,m,m',p,p',q,q'}= {} \frac{a}{r} \mca{ A }_{n,m,m',p,p',q,q'}
    U_{n,m,m'}(\epsilon) 
    \exp\ci \, \Theta_{n,m,m',p,p',q,q'}^{-} \;,
    \\
    \begin{split}
      & \mca{A}_{n,m,m',p,p',q,q'} = {}
      \frac{\mu'}{a'}
      \fracp*{a}{a'}^{n}
      (-1)^{m-m'} \frac{(n-m')!}{(n+m)!} 
      F_{n,m,p}  \left( I \right) F_{n,m',p'} \left( I' \right) 
      \\
      & \hspace{8em} \times Z_{q}^{n+1,n-2p}  \left( e \right) X_{q'}^{-(n+1),n-2p'}  \left( e' \right) \;,
    \end{split}
  \end{align}
\end{subequations}
or expressed in trigonometric form:%
\begin{subequations}
    \allowdisplaybreaks[1]
  \label{eq:R_moon_EO_trig_Fnmp}
  \begin{align}
    & \mca{R}_{\leftmoon} = {} \sum_{n \geq 2}^{} \,
    \sum_{m = 0}^{n} \,
    \sum_{m' = 0}^{n} \,
    \sum_{p = 0}^{n} \, 
    \sum_{p' = 0}^{n} \,
    \sum_{q = -n-1}^{n+1} \, 
    \sum_{q' = -\infty}^{+\infty} \, 
    \mca{R}_{n,m,m',p,p',q,q'} \;,
    \\
    \begin{split}
      & \mca{R}_{n,m,m',p,p',q,q'} =  {}
      \Delta_{0}^{m,m'} 
      \frac{a}{r} \mca{ A }_{n,m,m',p,p',q,q'}
      \\ & \hspace{8em} \times
      \left[ U_{n,m,m'} (\epsilon) \cos \Theta_{n,m,m',p,p',q,q'}^{-}
        \vphantom{\Big(} \right. \\ & \hspace{9em} \left. 
        + \, (-1)^{n-m'} U_{n,m,-m'} (\epsilon)  
        \cos\Theta_{n,m,m',p,p',q,q'}^{+} \right] \;,
    \end{split}
  \end{align}
\end{subequations}
where
\begin{subequations}
  \begin{alignat}{2}
    \Theta_{n,m,m',p,p',q,q'}^{\pm} 
    & = {} \Psi_{n,m,p,q} \pm \Psi'_{n,m',p',q'} \;,
    \\
    \Psi_{n,m,p,q} 
    & = {} qE + (n-2p) \omega + m \Omega \;,
    \\
    \Psi'_{n,m',p',q'} 
    & = {} q' M' + (n-2p') \omega' + m' \Omega' \;.
  \end{alignat}
\end{subequations}

Note that we have kept the term $a/r$ in factor, in order to anticipate the use of the relation \eqref{eq:20} in the following (Section \ref{sec-APPS}).

In case where the eccentricity of the third body is also high, we can introduce the Fourier series in multiples of the true anomaly $\nu$ (see Brumberg and Fukushima \cite{Brumberg_1994aa}) instead of the infinite Fourier series \eqref{eq:dev_fourier_M} defined by
\begin{equation}
  \label{eq:17}
  \fracp*{r'}{a'}^{-n} \exp\ci k'\nu' = 
  \sum_{q'=-\infty}^{+\infty} Y_{q'}^{-n,k'}(e') \exp\ci q'\nu' \;.
\end{equation}
The reason to use these Fourier series is that in this case, these series are finite for $-n\leq 0$. Indeed, the coefficients are null for $-n \geq \abs{q'-k'}$. We give in Annexe \ref{An-Ynms} their expression for this particular case.

Note that the average of the function $\Phi_{n,k}$ with respect to the mean anomaly $M$ over one period may be easily connected to coefficients $X$, $Y$ and $Z$:
\begin{align}
  & \mean*{ \Phi_{n,k}}_{M}  
  = {} \frac{1}{2 \pi} \int_{0}^{2 \pi} \Phi_{n,k} \diff{M}
  = {} X_{0}^{n,k} =  Z_{0}^{n+1,k} = \frac{1}{\eta} Y_{0}^{n+2,k}\;.
\end{align}

\section{Sun disturbing function} \label{sec-ESDF}

The general expression of the disturbing function for the Sun has been obtained by Kaula~\cite{Kaula_1962aa}. 
It can be obtained quickly by combining the expression of the disturbing function in spherical coordinates \eqref{eq:R_YYb} and the Fourier series of the spherical harmonics $Y_{n,m}$ given in~\eqref{eq:Gooding_Fnmk} for the perturbed and disturbing body. This yields,
\begin{subequations}
  \allowdisplaybreaks[1]
  \label{eq:R_sun_EO_exp_Fnmp}
  \begin{alignat}{2}
    \mca{R}_{\astrosun} & = {}
    \sum_{n \geq 2}^{} \,
    \sum_{m = -n}^{n} \,
    \sum_{p = 0}^{n} \,
    \sum_{p' = 0}^{n} \,
    \sum_{q = -n-1}^{n+1} \, 
    \sum_{q' = -\infty}^{+\infty} \, 
    \mca{R}_{n,m,p,p',q,q'} \;,
    \\
    \mca{R}_{n,m,p,p',q,q'} &= {} \frac{a}{r} \mca{A}_{n,m,p,p',q,q'}
    \exp\ci \, \Theta_{n,m,p,p',q,q'} \;,
    \\
    \begin{split}
      \mca{A}_{n,m,p,p',q,q'} & = {}
      \frac{\mu'}{a'}
      \left( \frac{a}{a'} \right)^{n}
      \frac{(n-m)!}{(n+m)!} 
      F_{n,m,p}  \left( I \right) F_{n,m,p'} \left( \epsilon \right)
      \\
      {} & \hspace{2em} \times
      Z_{q}^{n+1,n-2p}  \left( e \right) X_{q'}^{-(n+1),n-2p'}  \left( e' \right) \;,
    \end{split}
  \end{alignat}
\end{subequations}
with
\begin{subequations}
  \begin{alignat}{2}
    \Theta_{n,m,p,p',q,q'} 
    & = {} \Psi_{n,m,p,q} - \Psi'_{n,p',q'}  \;,
    \\
    \Psi_{n,m,p,q}  
    & = {} q E + (n-2p) \omega + m \Omega  \;,
    \\
    \Psi'_{n,p',q'} 
    & = {} q' M' + (n-2p') \omega' \;.
  \end{alignat}
\end{subequations}
or express in trigonometric form:
\begin{subequations}
  \label{eq:R_sun_HK_cos_Fnmp}
  \begin{align}
    \mca{R}_{\astrosun} = {} & \sum_{n \geq 2}^{} \,
    \sum_{m = 0}^{n} \,
    \sum_{p = 0}^{n} \, 
    \sum_{p' = 0}^{n} \,
    \sum_{q = -n-1}^{n+1} \, 
    \sum_{q' = -\infty}^{+\infty} \, 
    \mca{R}_{n,m,p,p',q,q'} \;,
    \\
    \mca{R}_{n,m,p,p',q,q'}  = {} &
    (2 - \delta_{0}^{m}) \frac{a}{r} 
    \mca{A}_{n,m,p,p',q,q'}
    \cos \Theta_{n,m,p,p',q,q'} \;.
  \end{align}
\end{subequations}

\section{Application of Lie Transform perturbation method} \label{sec-APPS}

The idea is to use a perturbative method based on the algorithm of the Lie Transform (see Deprit \cite{Deprit_1969aa}) in order to obtain an approximated analytical solution of the third-body problem. 
In this section, we give a method to treat the perturbed part  $\mca{H}_{1} = \mca{H}_{3c}$ of the Hamiltonian $\mca{H}$.

The third-body Hamiltonian $\mca{H}_{3c}$ is rewritten in order to isolate the secular, short-period and long-period parts:%
\begin{equation}
  \label{eq:2}
  \mca{H}_{3c} = {}
  \mca{H}_{3c} (y,Y,y',Y') = {} 
  {} \mca{H}_{3c,sec} + \mca{H}_{3c,sp} + \mca{H}_{3c,lp} \;.
\end{equation}
The secular part $\mca{H}_{3c,sec}$ is the part of $\mca{H}_{3c}$ that contains no term depending of any angular variable:%
\begin{subequations}
  \label{eq:134}
  \begin{align}
    \mca{H}_{3c,sec} 
    & = {} \mca{H}_{3c,sec} (\_,\_,Y,Y') \;,
    \\
    & = {} \lim_{T\to\infty}\frac{1}{T} \int_{0}^{T} \mca{H}_{3c} \, \diff{t} \;,
    \\
    & = \frac{1}{(2\pi)^{6}} {\int_{0}^{2\pi} \int_{0}^{2\pi} \cdots\!\int_{0}^{2\pi}} \mca{H}_{3c} \, \diff{l} \diff{l'} \diff{g}
    \diff{g'} \diff{h} \diff{h'} \;.
  \end{align}
\end{subequations}
Introduce the intermediate function%
\begin{align}
  \label{eq:132a}
  \mean*{\mca{H}_{3c}}_{l}  = {}
  \frac{1}{2\pi}\int_{0}^{2\pi} \mca{H}_{3c} \diff{l} = {}
  \frac{1}{2\pi}\int_{0}^{2\pi} \mca{H}_{3c} \frac{r}{a} \diff{E} \;.
\end{align}
The long-period perturbations $\mca{H}_{3c,lp}$ are computed by eliminating the secular terms inside~$\mean*{\mca{H}_{3c}}_{l}$:%
\begin{equation}
  \label{eq:5}
  \mca{H}_{3c,lp} = {}
  \mean*{\mca{H}_{3c}}_{l} - \mca{H}_{3c,sec} \;.
\end{equation}
The short-period perturbations $\mca{H}_{3c,sp}$ are computed by eliminating in $\mca{H}_{3c}$ all terms that do not depend on the fast variable~$l$:%
\begin{equation}
  \label{eq:4}
  \mca{H}_{3c,sp} = {}
  \mca{H}_{3c} - \mean*{\mca{H}_{3c}}_{l} \;,
\end{equation}
In practice, this sharing of $\mca{H}_{3c}$ is equivalent to an appropriate sorting of the indices in expression \eqref{eq:R_moon_HK_trig_Fnmp} or \eqref{eq:R_sun_HK_cos_Fnmp}.

We want to transform the Hamiltonian  $\mca{H}(y,y',Y,Y')$ to a new one $\mca{K}(y^{\star},y^{\prime},Y^{\star},Y^{\prime})$ by means of a generating function $\mca{W}$. 
The transformed Hamiltonian and the corresponding generator will be assumed expandable as power series of quantities having the same order of magnitude than $\Jd$:%
\begin{subequations}
  \begin{align}
    \label{eq:5}
    \mca{K} & = {}
    \mca{K}_{0} + \mca{K}_{1} + \gordre{2} \;,
    \\
    \mca{W} & = {}
    \mca{W}_{1} + \gordre{2} \;.
  \end{align}
\end{subequations}
Since the Hamiltonian $\mca{H}_{1}$ is time-dependent, we can solve \eqref{eq:sys_eq_Delaunay} using the first order time-dependent Lie Transform from \cite{Deprit_1969aa} as canonical perturbation method:%
\begin{subequations}
  \label{eq:Lie_upto_order1}
  \begin{alignat}{3}
    \label{eq:Lie_order0}
    \hspace{1em}\textbf{Order 0:}\quad & \mca{K}_{0} 
    && {} = \mca{H}_{0} \;,
    \\
    \label{eq:Lie_order1}
    \hspace{1em}\textbf{Order 1:}\quad & \mca{K}_{1} 
    && {} = \mca{H}_{3c}  + \cpoiss{\mca{H}_{0}}{\mca{W}_{1}} - \dpartial{\mca{W}_{1}}{t}
    \\ 
    {} & {} 
    && {} =
    \mca{H}_{3c}
    {-} \left( 
      \omega_{l} \dpartial{\mca{W}_{1}}{l} 
      + \omega_{g} \dpartial{\mca{W}_{1}}{g} 
      + \omega_{h} \dpartial{\mca{W}_{1}}{h} 
      + \dpartial{\mca{W}_{1}}{t}
    \right) \;,
  \end{alignat}
\end{subequations}
where $\cpoiss{\alpha}{\beta}$ is the Poisson brackets defined by%
\begin{equation}
  \label{eq:def_cpoiss}
  \cpoiss{\alpha}{\beta}_{y,Y} = {} \sum_{j=1}^{3} 
  \left(
    \dpartial{\alpha}{y_j} \dpartial{\beta}{Y_j}  - \dpartial{\alpha}{Y_j} \dpartial{\beta}{y_j}  
  \right)
  =  {-} \cpoiss{\beta}{\alpha}_{y,Y} \;,
\end{equation}
and the $\omega_{j}$ are the angular frequencies associated to $\mca{H}_{0}$: 
\begin{subequations}
  \begin{alignat}{4}
    \label{eq:67}
    \omega_{l} 
    & = {} \dot{l}_{0} 
    && = {} \dpartial{\mca{H}_{0}}{L} 
    && = {} n_{0} \left[ 1+6\gamma_{2} \, \eta \left( 1-3\cos^{2}I \right) \right] \;,
    \\
    \omega_{g} 
    & =  {} \dot{g}_{0} 
    && = {} \dpartial{\mca{H}_{0}}{G} 
    && = {} 6 \, \gamma_{2} \, n_{0} \left( 1-5\cos^{2}I \right) \;,
    \\
    \omega_{h} 
    & =  {} \dot{h}_{0} 
    && = {} \dpartial{\mca{H}_{0}}{H} 
    && = {} 12 \, \gamma_{2} \, n_{0} \cos I \;,
  \end{alignat}
\end{subequations}
with
\begin{equation}
  \gamma_{2}
  = - \frac{\Jd}{8 \eta^4} \fracp*{R_{\varoplus}}{a}^2
  = - \frac{\Jd}{8} \fracp*{\mu_{\varoplus} R_{\varoplus}}{G^2}^2 \;.
\end{equation}
We choose the new Hamiltonian $\mca{K}_{1}$ such that it does not depend on any angle variables: 
\begin{equation}
  \mca{K}_{1} = {} \mca{H}_{3c,sec} \;.
\end{equation}
Assuming that the angles $y'$ related to the third-body vary linearly with time and the momenta $Y'$ are constants (which is a good approximation), the $t$-partial derivative that appears in the homological equation \eqref{eq:Lie_order0} can be expressed as:%
\begin{equation}
  \label{eq:dW1sdt}
  \dpartial{\mca{W}_{1}}{t} = {} 
  \omega_{l'} \dpartial{\mca{W}_{1}}{l'} 
  + \omega_{g'} \dpartial{\mca{W}_{1}}{g'} 
  + \omega_{h'} \dpartial{\mca{W}_{1}}{h'} \;.
\end{equation}
where $\omega_{l'} = \dot{l}_{sec}^{\prime}$, $\omega_{g'} = \dot{g}_{sec}^{\prime}$ and $\omega_{h'} = \dot{h}_{sec}^{\prime}$.

Substituting \eqref{eq:dW1sdt} in \eqref{eq:Lie_order1}, we obtain the PDE:%
\begin{equation}
  \label{eq:homologic_1}
  \dpartial{\mca{W}_{1}}{l} + \sum_{j=1}^5 \beta_j \dpartial{\mca{W}_{1}\textbf{}}{\alpha_j} = {}
  \frac{1}{\omega_{l}} \left( 
    \mca{H}_{3c} -\mca{K}_{1}
  \right) = {}
  \frac{1}{\omega_{l}} \left( 
    \mca{H}_{3c,sp} +\mca{H}_{3c,lp}
  \right)
  \;,
\end{equation}
where
\begin{subequations}
  \begin{align}
    \label{eq:alpha_j}
    \alpha_{j}  
    & = \left\{ g, h, l', g', h' \right\} \;,
    \\
    \label{eq:omega_j}
    \omega_{j} 
    & = \left\{ \omega_{g}, \omega_{h}, \omega_{l'}, \omega_{g'}, \omega_{h'} \right\} \;,
    \\
    \label{eq:beta_j}
    \beta_{j} 
    & = {} \dfrac{\omega_{j}}{\omega_{l}} \ll 1\;.
  \end{align}
\end{subequations}
As we can note, we have introduced in \eqref{eq:homologic_1} a small parameter $\beta_{j}$, which is the ratio between the low angular frequency and $\omega_{l}$. 
Since the fastest long-period is $2\pi/\omega_{l'}$ (about~$28$ days for the Moon) and supposing that the satellite orbital period for a highly elliptic orbit can reach 1--2 days, $\beta_{j}$ can not exceed~${1/15}$.

In order to solve the PDE, we seek a solution in the form 
\begin{align}
  \label{eq:118}
  \mca{W}_{1} = {} \mca{W}_{1,lp}(\_,g,h,y',Y,Y') + \mca{W}_{1,sp}(l,g,h,y',Y,Y') 
  \;,
\end{align}
such that
\begin{subequations}
  \begin{empheq}[left=\empheqlbrace]{alignat=1}
    \label{eq:EDP_a} 
    \sum_{j=1}^5 \beta_j\dpartial{\mca{W}_{1,lp}}{\alpha_j} 
    & = {} \frac{1}{\omega_{l}} \mca{H}_{3c,lp}  \;,
    \\
    \label{eq:EDPb} 
    \dpartial{\mca{W}_{1,sp}}{l}+\sum_{j=1}^5 \beta_j\dpartial{\mca{W}_{1,sp}}{\alpha_j} 
    & = {} \frac{1}{\omega_{l}} \mca{H}_{3c,sp}  
    \;.
  \end{empheq}
\end{subequations}

The solution for the generator of the long-periods $\mca{W}_{1,lp}$ is straightforward, while the solution for the generator of short-periods $\mca{W}_{1,sp}$ is more technical.
To bypass this difficulty, we solve the PDE \eqref{eq:EDPb} by means of a recursive process which may be more suitable for an analytical theory. 

To solve \eqref{eq:EDPb}, we need to integrate all terms with respect to the mean anomaly $l$. Since our disturbing function is express in term of the eccentric anomaly $E$ we can use the relation \eqref{eq:20} that connects the eccentric anomaly $E$ and the mean anomaly $l$. The PDE rewrites
\begin{equation}
  \label{eq:89}
  \dpartial{\mca{W}_{1,sp}}{E} +\sum_{j=1}^5 \frac{r}{a} \beta_j \dpartial{\mca{W}_{1,sp}}{\alpha_j} = {} 
  \frac{1}{\omega_{l}}\frac{r}{a} \mca{H}_{3c,sp} \;.
\end{equation}
We note that we have in factor of $\mca{H}_{3c,sp}$ the ratio $r/a$. This term will simplify due to the fact that we have anticipated this factor in the development of the disturbing function \eqref{eq:R_moon_EO_exp_Fnmp} and \eqref{eq:R_sun_EO_exp_Fnmp}.

Given that $\beta_{j} \ll 1$, we can assume that the generator $\mca{W}_{1,sp}$ is expandable in power series of the quantity $\beta_{j}$:%
\begin{equation}
  \label{eq:Taylor_W1sp}
  \mca{W}_{1,sp} 
  = {} \mca{W}_{1,sp}^{^(0)} + \sum_{\sigma \ge 1}^{} \mca{W}_{1,sp}^{(\sigma)} \;.
\end{equation}
Inserting this series in \eqref{eq:89}, the generator $\mca{W}_{1,sp}$ can be recursively determined by using the relations%
\begin{subequations}
  \label{eq:91}
  \begin{empheq}[left=\empheqlbrace]{alignat=1}
    \label{eq:91a} 
    \dpartial{\mca{W}_{1,sp}^{(0)}}{E} & = {} 
    \frac{1}{\omega_{l}} \frac{r}{a} \mca{H}_{3c,sp} \;,
    \\
    \label{eq:91b} 
    \dpartial{\mca{W}_{1,sp}^{(\sigma+1)}}{E} & = {} 
    {-} \sum_{j=1}^5 \frac{r}{a} \beta_j \dpartial{\mca{1,W}_{sp}^{(\sigma)}}{\alpha_j} \;, \quad \sigma \geq 0
    \;.
  \end{empheq}
\end{subequations}
The order zero is considered as the initial guess and the order $(\sigma+1)$ as a correction of the solution of order $\sigma$.
Also, we impose that the mean value of the generator $\mca{W}_{1,sp}^{(\sigma)}$ over the mean anomaly $l$ is zero: $\mean*{\mca{W}_{1,sp}}_{l} = 0$. This can be realized by adding a constant $C^{(\sigma)}$ independent of the eccentric anomaly.



\section{Conclusion}




The construction of an analytical theory of the third-body perturbations in case of highly elliptical orbits is facing several difficulties. 
On the one hand, the Fourier series in term of the mean anomaly converge slowly, on the other hand the disturbing function is time dependent. 
Each of these difficulties can be solved separately with more-or-less classical methods. 
Concerning the first issue, it is already known that the Fourier series in multiple of eccentric anomaly are finite series. Their use in an analytical theory is less simple than classical series in multiple of the mean anomaly, but remains tractable. The time dependence is not a great difficulty, only a complication: after having introduced the appropriate (time linear) angular variables in the disturbing function, these variables must be taken into account in the PDE to solve during the construction of the theory. 

Combining the two problems (expansion in terms of the eccentric anomaly and time dependence) in the same theory is a more serious issue. 
In particular, solving the PDE \eqref{eq:homologic_1} in order to express the generator of the canonical transformation is not trivial. 
In this paper we propose two recipes to progress towards a solution: 
\begin{itemize}
\item an appropriate expansion of the disturbing function involving the Fourier series with respect to the eccentric anomaly; more precisely, this is not a complete expansion since we keep the ratio $a/r$ in factor for the needs of the theory;
\item the generator, solution of the PDE \eqref{eq:EDPb} is constructed by means of an iterative process, which is equivalent to a development of a generator in power series of small ratio of slow to fast angular frequencies.
\end{itemize}
The inclusion of this iterative process in a fully automatic and recursive algorithm is in progress. 
This allows to get a very compact solution using special functions. The main advantage is that the degree of approximation of the solution (e.g. the truncation of the development in spherical harmonics~\eqref{eq:R_YYb} and the number of iterations in the resolution of~\eqref{eq:91b}) are chosen by the user as needed and not fixed once for all when constructing the theory.

As said in Section 1, this paper does not deal with the perturbation due to $\Jd$. In principle this is a well-known problem having received numerous satisfactory solutions. However, we have included the $\Jd$ secular terms in the main part of the Hamiltonian $\mca{H}_0$. The advantage is that this Hamiltonian is not degenerate (it contains the three momenta~$L$, $G$ and $H$), and consequently, the homological equation \eqref{eq:Lie_order1} allows to choose a new Hamiltonian free of any angular variable. In return, the most classical theories (e.g. \cite{Brouwer_1959aa}, \cite{Abad_2008aa}, \cite{San-Juan_2011aa}) for which the whole $\Jd$ perturbation is relegated in $\mca{H}_1$ are no longer directly usable. We have to construct another solution taking into account the new sharing of the $\Jd$ disturbing function. This work is in progress.


\appendix

\section{Second proof of the trigonometric formulation}\label{An-Method_2}

Let us define the spherical harmonics as
\begin{subequations}
  \begin{alignat}{3}
    Y_{n,m} 
    & = {} Y_{n,m}(\delta, \alpha) 
    && = {} C_{n,m}(\delta, \alpha) + \ci S_{n,m}(\delta, \alpha) \;,
    \label{eq:Ynm_sat_eq}
    \\
    Y^{\prime\ast}_{n,m} 
    & = {} Y^{\ast}_{n,m}(\delta', \alpha') 
    && = {} C'_{n,m}(\delta', \alpha') - \ci S'_{n,m}(\delta', \alpha') \;.
    \label{eq:Ynm_3c_eq}
  \end{alignat}
\end{subequations}
The disturbing function \eqref{eq:R_YYb} can be expanded in trigonometric form as
\begin{equation}
  \label{eq:R_YYb_trig}
  \mca{R}_{3c} = {} 
  \frac{\mu'}{r'} 
  \sum_{n \geq 2}^{} \sum_{m = 0}^{n} 
  (2 - \delta_{0}^{m})
  \frac{(n-m)!}{(n+m)!} \fracp*{r}{r'}^n 
  \Re \left\lbrace Y_{n,m} Y^{\prime\ast}_{n,m} \right\rbrace  \;,
\end{equation}
where
\begin{equation}
  \label{eq:Re_YYb}
  \Re\left\lbrace{Y_{n,m} Y^{\prime\ast}_{n,m}}\right\rbrace = C_{n,m} C'_{n,m} + S_{n,m} S'_{n,m} \;.
\end{equation}
At first, we express \eqref{eq:Re_YYb} in terms osculating orbital elements of the satellite with the help of the transformation \eqref{eq:Ynm_Sat_eo2eq_Fnmp}. We get
\begin{equation}
  \begin{split}
    \Re \left\{ Y_{n,m} Y^{\prime\ast}_{n,m} \right\} = {} &
    \ci^{n-m} \sum_{p = 0}^{n} F_{n,m,p} (I)
    \left\{ 
      \left[ 
        \begin{array}{r}
          C'_{n,m} \\ - \ci S'_{n,m}
        \end{array}
      \right]_{(n-m)_{\mathrm{odd}}}^{(n-m)_{\mathrm{even}}}
      \cos \Psi_{n,m,p}
    \right. \\ {} & {} \left.
      \hspace{76pt} + {}
      \left[ 
        \begin{array}{r}
          S'_{n,m} \\ \ci C'_{n,m}
        \end{array}
      \right]_{(n-m)_{\mathrm{odd}}}^{(n-m)_{\mathrm{even}}}
      \sin \Psi_{n,m,p}
    \right\} \;.
  \end{split}
\end{equation}

We express then the elements $C'_{n,m}$ and $S'_{n,m}$ in terms osculating orbital elements of the Moon.

Let be 
\begin{align}
  & U_{n,m,m'}^c = {} \frac{1}{2} \left( U_{n,m,m'} + U_{n,m,-m'} \right) \;,
  \\
  & U_{n,m,m'}^s = {} \frac{1}{2} \left( U_{n,m,m'} - U_{n,m,-m'} \right) \;,
\end{align}
with the $U$-functions defined in \eqref{eq:Wigner_U}. 

Splitting the sum over $m'$ of the transformation \eqref{eq:Ynm_Moon_ec2eq} into two parts such that $m' \geq 0$ and using the relation \eqref{eq:Ynm_m_neg}, it follows that
\begin{align}
  \begin{split}
    Y_{n,m} 
    & = \sum_{m' = 0}^{n} \ci^{m-m'} 
    {} (2-\delta_{0}^{m'}) \frac{(n-m')!}{(n-m)!} P_{n,m'}(\sin \beta')
    \\ & \hspace{2em} \times
    \left[ U_{n,m,m'}^c \cos m' \lambda' + \ci U_{n,m,m'}^s \sin m' \lambda' \right] \;.
  \end{split}
\end{align}
Contrary to Giacaglia \cite{Giacaglia_1974aa} and Lane \cite{Lane_1989aa}, the complex power $\ci^{m'}$ is not absorbed into the trigonometric terms. 
This permits to reduce the complexity of calculations naturally by setting constraints on the parity of the couple $n-m$ and $m-m'$.
Thus, the real part of~$Y'_{n,m}(\delta', \alpha')$ is given by
\begin{equation}
  C'_{n,m} =
  \sum_{m' = 0}^{n} \ci^{m-m'} (2-\delta_{0}^{m'}) \frac{(n-m')!}{(n-m)!} 
  \left[ 
    \begin{array}{r}
      U_{n,m,m'}^c \, C''_{n,m'} \\ 
      \ci U_{n,m,m'}^s \, S''_{n,m'}
    \end{array}
  \right]_{(m-m')_{\mathrm{odd}}}^{(m-m')_{\mathrm{even}}}
  \label{eq:Cp} \;,
\end{equation}
and the imaginary part by
\begin{equation}
  S'_{n,m} =
  \sum_{m' = 0}^{n} \ci^{m-m'} (2-\delta_{0}^{m'}) \frac{(n-m')!}{(n-m)!} 
  \left[ 
    \begin{array}{r}
      U_{n,m,m'}^s \, S''_{n,m'}  \\ -\ci U_{n,m,m'}^c \, C''_{n,m'}
    \end{array}
  \right]_{(m-m')_{\mathrm{odd}}}^{(m-m')_{\mathrm{even}}} \;,
  \label{eq:Sp}
\end{equation}
where $C''_{n,m'} = \Re\left\lbrace Y_{n,m'}(\beta',\lambda') \right\rbrace$ and $S''_{n,m'} = \Im\left\lbrace Y_{n,m'}(\beta',\lambda')\right\rbrace$.

Converting these coefficients in terms of osculating orbital elements referred in the ecliptic plane thanks to the transformation \eqref{eq:Ynm_Moon_eo2ec_Fnmp}, we find
\begin{subequations}
  \begin{align}
    C''_{n,m'} & = \ci^{n-m'}	\sum_{p' = 0}^{n}
    \left[ 
      \begin{array}{r}
        \cos \Psi'_{n,m',p'} \\ \ci \sin \Psi'_{n,m',p'}
      \end{array}
    \right]_{(n-m')_{\mathrm{odd}}}^{(n-m')_{\mathrm{even}}} \;,
    \label{eq:Yc_EO}
    \\
    S''_{n,m'} & = \ci^{n-m'}	\sum_{p' = 0}^{n}
    \left[ 
      \begin{array}{r}
        \sin \Psi'_{n,m',p'} \\ -\ci \cos \Psi'_{n,m',p'}
      \end{array}
    \right]_{(n-m')_{\mathrm{odd}}}^{(n-m')_{\mathrm{even}}} \;.
    \label{eq:Ys_EO}
  \end{align}
  \label{eq:Ypp}
\end{subequations}

Before replacing anything, it is interesting to take a look on the possible parities between the values of~$n-m$, $m-m'$ and $n-m'$ (see Table \ref{tab_parity}). Indeed, if we fix the parity of~$n-m$ and $m-m'$, we constrain the parity of $n-m'$.

\begin{table}[h!]
  \centering
  \begin{tabular}{ *{4}{c} }    
    \toprule
    Case &  $n-m$  &  $m-m'$  &  $n-m'$ \\
    \cmidrule( r){1-1}
    \cmidrule(  ){2-3}
    \cmidrule(l ){4-4}
    (a)  &  even  &  even  &  even \\ 
    (b)  &  even  &  odd   &  odd  \\ 
    (c)  &  odd   &  even  &  odd  \\ 
    (d)  &  odd   &  odd   &  even \\ 
    \bottomrule
  \end{tabular}
  
  \begin{minipage}{0.40\textwidth}
    \centering
    \caption{Possible parities.}
    \label{tab_parity}
  \end{minipage}
\end{table}

Consider 
\begin{equation}
  \mca{R} = {}
  \sum_{n \geq 2}^{} \,
  \sum_{m = 0}^{n} \,
  \sum_{m' = 0}^{n} \,
  \sum_{p = 0}^{n} \,
  \sum_{p' = 0}^{n}
  \mca{R}_{n,m,m',p,p'}  \;.
\end{equation}
Substituting now the relations \eqref{eq:Ypp} into \eqref{eq:Cp} and \eqref{eq:Sp}, we find according the cases listed in Table~\ref{tab_parity}:%
{
  \begin{subequations}
    \begin{alignat}{5}
      & \mca{R}_{n,m,m',p,p'} 
      && {} = (-1)^{n-m'} \mcat{R}_{n,m,m',p,p'} \times \dots \nonumber
      \\
      \mbox{\hspace{2ex}\textbullet\hspace{1ex} (a)} 
      & 
      && \quad \left[ \hgm U_{n,m,m'} C_{n,m,m',p,p',q,q'}^{-} + U_{n,m,-m'}
        C_{n,m,m',p,p',q,q'}^{+} \right] \;,
      \\
      \mbox{\hspace{2ex}\textbullet\hspace{1ex} (b)}
      &
      && \quad \left[ \hgm U_{n,m,m'} C_{n,m,m',p,p',q,q'}^{-} - U_{n,m,-m'}
        C_{n,m,m',p,p',q,q'}^{+} \right] \;,
      \\
      \mbox{\hspace{2ex}\textbullet\hspace{1ex} (c)}
      & 
      && \quad \left[ -U_{n,m,m'} C_{n,m,m',p,p',q,q'}^{-} + U_{n,m,-m'} C_{n,m,m',p,p',q,q'}^{+}
      \right] \;,
      \\
      \mbox{\hspace{2ex}\textbullet\hspace{1ex} (d)} 
      & 
      && \quad\left[ -U_{n,m,m'} C_{n,m,m',p,p',q,q'}^{-} - U_{n,m,-m'} C_{n,m,m',p,p',q,q'}^{+}
      \right] \;,
    \end{alignat}
  \end{subequations}
}
with
\begin{align}
  & \mcat{R}_{n,m,m',p,p'} = {} 
  \Delta_{0}^{m,m'} \frac{(n-m')!}{(n+m)!} 
  \frac{\mu'}{r'} \fracp*{r}{r'}^{n} F_{n,m,p}(I) F_{n,m',p'}(I')  \;,
  \\
  & C_{n,m,m',p,p',q,q'}^{\pm} = {} \cos \left( \Psi_{n,m,p,q} \pm \Psi'_{n,m',p',q'} \right)  \;.
\end{align}
Finally, since the sign of $U_{n,m,m'}$ depends on the parity of $n-m$ and those of $U_{n,m,-m'}$ of the parity of $m-m'$, we can combine the four relations into only one leading to the relation~\eqref{eq:R_moon_HK_trig_Fnmp}.

\section{Hansen-like coefficients $Z_{s}^{n,m}$ and $Y_{s}^{n,m}$}\label{An-Znms}

Contrary to the classical Hansen coefficients $X_{s}^{n,m} $, the coefficients $Y_{s}^{n,m} $ and~$Z_{s}^{n,m} $
(or equivalents) are not widely used and, to our knowledge, only a few authors have worked about their
analytical expression.  
We can quote (Brown \cite{Brown_1933aa}, p. 68-71; Brumberg and Fukushima~\cite{Brumberg_1994aa}; Da Silva Fernandes \cite{Da-Silva-Fernandes_1996aa} and or more recently Laskar \cite{Laskar_2005aa}.
\\
In this section, we present a simple method to obtain the expressions of $Y_{s}^{n,m}$ for $n \leq 0$ and~$Z_{s}^{n,m}$ for $n \geq 0$, which corresponds to our need in this paper.

\subsection{Coefficients $Z_{s}^{n,m}$}
\label{sec:ggf}

The Hansen-like coefficients $Z_{s}^{n,m} $ defined in \cite{Brumberg_1994aa} as
\begin{equation}
  \Phi_{n,m} = {} 
  \sum_{s=-\infty}^{+\infty} Z_{s}^{n,m} (e) \exp\ci sE \;,
  \label{eq:dev_fourier_E}
\end{equation}
depend on the eccentricity $e$ and admit an integral representation:
\begin{align}
  Z_s^{n,m} &= {} \frac{1}{2\pi} \int\limits_{0}^{2\pi} \Psi_s^{n,m} \diff{E} \;,
  \\ 
  \Psi_s^{n,m} & = {} \left(\frac{r}{a} \right)^{n} \sigma^m \zeta^{-s} \;.
  \label{eq:Znms_int_E}
\end{align}
with $\sigma = \exp \ci \nu$  and $\zeta = \exp \ci E$ .

Due to the fact that $\nu$ is an odd function of $E$, if we change $m$ by $-m$ and $s$ by $-s$ in~\eqref{eq:dev_fourier_E}, we get directly the symmetry:
\begin{equation}
  Z_{-s}^{n,-m} = Z_{s}^{n,m} \;,
  \label{eq:Znms_sym}
\end{equation}
as in the case of  the classical Hansen coefficients $X_{s}^{n,m}$. 

According to Tisserand  \cite[p. 251-252]{Tisserand_1889aa}, one can express $r/a$ and $\sigma$ in terms of $\zeta$ as 
\begin{align}
  \frac{r}{a} &= \frac{1}{1+\beta^2} (1-\beta \zeta^{-1}) (1-\beta \zeta) \;, \\
  \sigma &= \zeta ( 1-\beta{\zeta}^{-1} ) \left( 1-\beta\zeta \right)^{-1} \;,
\end{align}
with $\beta=\dfrac{e}{1+\sqrt{1-e^2}}$. 

Consequently
\begin{equation}
  \Psi_s^{n,m}(\zeta)= \frac{1}{(1+\beta^2)^{n}} (1-\beta \zeta)^{n-m}  (1-\beta \zeta^{-1})^{n+m} \zeta^{m-s} .
\end{equation}
Since $0 \leq \beta < 1$ for elliptic motion, using the Taylor expansions 
\begin{equation}
  \begin{split}
    & (1-\beta \zeta)^{n-m} = \sum_{p=0}^{+\infty} \binom{n-m}{p} (-\beta)^p \zeta^p  \;, \\
    & (1-\beta \zeta^{-1})^{n+m}= \sum_{q=0}^{+\infty} \binom{n+m}{q} (-\beta)^q \zeta^{-q} \;,
  \end{split}
\end{equation}
we get
\begin{equation}
  \label{eq:14}
  \Psi_s^{n,m}  
  = \frac{1}{(1+\beta^2)^{n}} \sum_{q=0}^{+\infty} \sum_{p=0}^{+\infty}
  \binom{n-m}{p} \binom{n+m}{q} (-\beta)^{p+q} \zeta^{m-s+p-q} 
  \;,
\end{equation}
which can be rewritten in a power series of $\zeta$ as
\begin{equation}
  \label{eq:15}
  \Psi_s^{n,m}  = \sum_{r=-\infty}^{+\infty} \alpha_{s,r}^{n,m} \zeta^r \;.
\end{equation}
Restricting to $n \geq 0$ and inserting \eqref{eq:15} into \eqref{eq:Znms_int_E}, we get 
\begin{equation}
  \label{eq:16}
  Z_{s}^{n,m} = \alpha_{s,0}^{n,m} = {} 
  Z_{-s}^{n,-m} = (-1)^{m-s} \frac{\beta^{m-s}}{(1+\beta^2)^{n}} 
  \sum_{q = q_{min}}^{q_{max}} 
  \binom{n-m}{q} \binom{n+m}{q+m-s} \beta^{2q} 
  \;.
\end{equation}
Here $q$ takes all values making the binomial coefficients non-zero, i.e. from
$\linebreak[0]{q_{min} = \max(0,s-m)}$ to $q_{max} = n+s$.
Additionally, if $0 \leq \abs{m} \leq n$, the upper limit of the sum is given by $q_{max} = \min(n-m,n+s)$ and for $
|s| >n$ we have $ Z_{s}^{n,m} =0$. 
Thus, the representation of the $Z_{s}^{n,m}$ is finite.

More relations and efficient algorithms to compute the $ Z_{s}^{n,m}$ coefficients are given in Lion and Métris \cite{Lion_2013aa}.

\subsection{Coefficients $Y_{s}^{n,m}$}
\label{An-Ynms}


The coefficients $Y_{s}^{n,m} (e)$ are defined by
\begin{equation}
  \left( \frac{r}{a} \right)^{n} \exp \ci m\nu 
  = \sum_{s=-\infty}^{+\infty} Y_{s}^{n,m} (e) \exp \ci s\nu \;.
  \label{eq:dev_fourier_v}
\end{equation}
Setting
\begin{equation}
  \Psi_s^{n,m} = \left(\frac{r}{a} \right)^{n} \sigma^{m-s} \;,
  \label{eq:Psi}
\end{equation}
each coefficient admits an integral representation
\begin{equation}
  Y_s^{n,m} (e) = \frac{1}{2\pi} \int\limits_{0}^{2\pi} \Psi_s^{n,m} \mathrm{d}\nu \;,
  \label{eq:int_Y}
\end{equation}
and the symmetry 
\begin{equation}
  Y_{s}^{n,m} = Y_{-s}^{n,-m} = Y_{0}^{n,m-s} = Y_{s-m}^{n,0} \;.
  \label{eq:Ynms_sym}
\end{equation}
Repeating the process used for  $Z_{s}^{n,m}$, we expand $\Psi_s^{n,m}$ in power of $\sigma$ (instead of $\zeta$)  as follows
\begin{equation}
  \Psi_s^{n,m} = \sum_{r=-\infty}^{+\infty} \alpha_{s,r}^{n,m} \sigma^m \;.
\end{equation}
Knowing that
\begin{equation}
  \frac{a}{r} = \frac{1+\beta^2}{(1-\beta^2)^2} (1+\beta \sigma^{-1}) (1+\beta \sigma) \;,
\end{equation}
and restricting to the case $n \leq 0$ we find
\begin{subequations}
  \begin{align}
    Y_{s}^{n,m} 
    & = \alpha_{s,0}^{n,m} = Y_{-s}^{n,-m} = Y_{s-m}^{n,0} \;,
    \\
    & = \beta^{m-s} \frac{(1-\beta^2)^{2n}}{(1+\beta^2)^n} 
    \sum_{q=q_{min}}^{q_{max}} \binom{-n}{q+m-s} \binom{-n}{q} \beta^{2q} 
    \;,
    \label{eq:Ynms}
  \end{align}
\end{subequations}
where $q$ takes the values making the binomial coefficients nonzero, i.e. 
${q_{min} =\max(0,m-s)}$ and ${q_{max} =\min(-n,-n-m+s)}$.
%
%
It is easy to show that for $\mid m-s\mid>-n$ we have $ Y_{s}^{n,m} = 0$, which proves that expansions of the form
(\ref{eq:dev_fourier_v}) admits a closed-form representation for~${-n \geq 0}$.


\end{document}